\documentclass[aps,prl,reprint,superscriptaddress]{revtex4-2}

\usepackage{physics}
\usepackage[T1]{fontenc} 
\usepackage[utf8x]{inputenc}
\usepackage{hyperref}
\usepackage{subcaption}
\usepackage{pgfplots}
\usetikzlibrary{angles,quotes}
\usepackage{soul}
 
\setstcolor{red}

\begin{document}


\title{Vortex structure deformation of rotating Lifshitz holographic superconductors}


\author{Jhony A. Herrera-Mendoza}
\email{jherrera@ifuap.buap.mx}
\affiliation{Instituto de F\'{i}sica, Benem\'{e}rita Universidad Aut\'{o}noma de Puebla, Edificio IF-1, Ciudad Universitaria, CP 72570, Puebla, M\'{e}xico.}

\author{Daniel F. Higuita-Borja}
\email{dhiguita@ifuap.buap.mx}
\affiliation{Instituto de F\'{i}sica, Benem\'{e}rita Universidad Aut\'{o}noma de Puebla, Edificio IF-1, Ciudad Universitaria, CP 72570, Puebla, M\'{e}xico.}

\author{Julio A. M\'{e}ndez-Zavaleta}
\email{julmendez@uv.mx}
\affiliation{Facultad de F\'{i}sica, Universidad Veracruzana, Paseo No. 112, Desarrollo Habitacional Nuevo Xalapa, C.P. 91097, Xalapa-Enr\'iquez, M\'{e}xico.}

\author{Alfredo Herrera-Aguilar}
\email{aherrera@ifuap.buap.mx}
\affiliation{Instituto de F\'{i}sica, Benem\'{e}rita Universidad Aut\'{o}noma de Puebla, Edificio IF-1, Ciudad Universitaria, CP 72570, Puebla, M\'{e}xico.}

\author{Felipe P\'{e}rez-Rodr\'{i}guez}
\email{fperez@ifuap.buap.mx}
\affiliation{Instituto de F\'{i}sica, Benem\'{e}rita Universidad Aut\'{o}noma de Puebla, Edificio IF-1, Ciudad Universitaria, CP 72570, Puebla, M\'{e}xico.}

\date{\today}

\begin{abstract}
We construct a type-II holographic superconductor from an anisotropic rotating background metric.
We study the effects of the magnetic field on the vortex structure, including continuous deformations from triangular to square lattices or vice versa. Our holographic model reproduces known experimental vortex lattice deformations and the increment of the vortex population by increasing the external magnetic field.
\end{abstract}


\maketitle

\emph{Introduction.}---
Nowadays, the gauge/gravity duality has become an important tool to describe different physical phenomena. Since the original proposal, \emph{AdS/CFT correspondence} \cite{Maldacena:1997re}, the framework has been extended to explore different phenomena such as quark-gluon plasma \cite{Casalderrey-Solana:2011dxg}, superfluidity \cite{Herzog:2008he}, and superconductivity \cite{Hartnoll:2008vx,Hartnoll:2008kx,Hartnoll:2009sz}. The seminal works in superconductivity qualitatively reproduce the condensation operator for different boundary conditions. 
 Ever since, major progress has been made to extend the holographic description of the phenomena. The vortex structure was explored for the first time in \cite{Montull:2009fe}, while a clearer analytical procedure was implemented in \cite{Maeda:2009vf}, reproducing the Abrikosov vortex structure \cite{ABRIKOSOV1957199}. 
 A degree of anisotropy has been also considered via Lifshitz black holes in arbitrary dimension \cite{ZHAO2014438}. For both the condensate and the magnetic fields, a good agreement between the semianalytical matching method and the numerical methods has been reported.
  The holographic superconductor for rotating AdS-type spacetimes was constructed in \cite{Sonner:2009fk,Lin:2014tza} where the introduction of a magnetic potential component $A_\varphi$ results mandatory to support the configuration and make compatible the boundary conditions. A noteworthy re-visit to the problem is presented in \cite{Srivastav:2019ixc}, where analytical methods were pursued instead of a numerical approach.
   Several details of the Lifshitz holography were collectively worked out in \cite{Guo:2014wca,Natsuume:2018yrg}, including both the condensate operator and the vortex structure. To the best of the authors' knowledge, this is the first report on vortex lattice deformations in type-II superconductors, realized from a rotating holographic description. In  previous works, the consideration of anisotropy and angular momentum achieve solely a modification of the lattice scale. 

\emph{The setup.}---
Our starting point is the 5-dimensional anisotropic rotating  metric  
\begin{align}\label{eq:Cov_Ansatz5D}
ds^2= & -\left(\dfrac{r}{\ell}\right)^{2z}f(r)\left(\Xi dt-ad\phi\right)^2+\dfrac{r^2}{\ell^4}\left(a dt-\Xi \ell^2 d\phi\right)^2\nonumber\\
& +\dfrac{dr^2}{\left(\dfrac{r}{\ell}\right)^2f(r)}+\left(\dfrac{r}{\ell}\right)^2d\vec{y}{}^2, 
\end{align}
where $\Xi=\sqrt{1+a^2/\ell^2}$, $d\vec{y}{}^2=dx^2+dy^2$, $f(r)=1-\left(\dfrac{r_h}{r}\right)^{z+3}$, $z$ is the critical exponent, $a$ is the rotation parameter, $\ell$ is the Lifshitz radius and $r_h$ is the horizon radius. 
Spacetime \eqref{eq:Cov_Ansatz5D} stems from the Einstein-Maxwell-dilaton gravity 
\begin{equation}\label{eq:EMD}
S=\int{d^5x\sqrt{-g}\left[\dfrac{R-2\lambda}{2\kappa}-\dfrac{1}{2}\left(\nabla_\mu\varphi\right)^2-\dfrac{1}{4}e^{-b\varphi}\mathcal{F}_{\mu\nu}\mathcal{F}^{\mu\nu}\right]},
\end{equation}
for suitable Maxwell and dilaton fields $\mathcal{A}_\mu$ and $\varphi$ respectively. The metric describes a 5-dimensional black hole, and it falls into a particular case of the solutions reported in \cite{Herrera-Aguilar:2021top}. The black hole temperature, a fundamental quantity for the upcoming analysis, is given by
\begin{equation}\label{eq:BBTemperature}
 T=\dfrac{1}{4\pi}\dfrac{(z+3)r_h^z}{\ell^{z+1}\Xi}.
\end{equation}
The holographic model is well described by the action
\begin{equation}\label{eq:EMD}
S_\text{m}=\int{d^5x\sqrt{-g}\left[\dfrac{1}{4}F_{\mu\nu}F^{\mu\nu}+\dfrac{1}{\ell^2}\left(|D_\mu\Psi|^2+m^2|\Psi|^2\right)\right]},
\end{equation}
with equations of motion
\begin{align}\label{eq:GenFieldEqns}
 \nabla_\mu F^{\mu\nu}-\dfrac{1}{\ell^2}\left(2A^\nu|\Psi|^2-i\Psi\nabla^\nu\overline{\Psi}+i\overline{\Psi}\nabla^\nu\Psi\right) & =0,\nonumber\\
 \nabla^{\mu}\nabla_{\mu}\Psi-2iA^{\mu}\nabla_{\mu}\Psi-i\Psi\nabla_{\mu}A^{\mu}-\left(m^2+A_{\mu}A^{\mu}\right)\Psi & =0,
\end{align}
where $D_\mu\equiv\nabla_\mu-iA_\mu$ is the gauge covariant derivative. For simplicity, we adopt throughout the rest of the text the new coordinate patch $(t, \phi, u, x, y)$, with $u = r_h/r$ and where the boundary and horizon are located at $u=0$ and $u=1$ respectively. 

\emph{The effect of an external magnetic field.}---
In this section, we investigate the effects of placing the type II holographic s-wave superconductor under the influence of an external magnetic field. It is well known that type II superconductors are characterized by manifesting two critical magnetic fields, a lower one  $B_{c1}$ and an upper one $B_{c2}$. The vorticity phenomenon appears when the external magnetic field surpasses the lower critical value $B>B_{c1}$. However, when the external magnetic field approaches the upper critical value $B_{c2}$ from below, a vortex lattice is prone to be formed. On the other hand, by decreasing the external magnetic field in the direction perpendicular to the boundary ($u=0$) while keeping the temperature and chemical potential fixed, it can be seen that the scalar field $\Psi$ begins to condensate below the upper critical value $B_{c2}$.

To construct the vortex lattice, we will use a generalized version of the procedure introduced in \cite{Maeda:2009vf}. First, we need to solve system in equation \eqref{eq:GenFieldEqns} near the upper critical magnetic field. With this aim, we perform a series expansion of the scalar and gauge fields on the parameter $\epsilon = (B_{c2}-B)/B$:
\begin{subequations}\label{eq:epsilon_expan}
\begin{align} 
&\Psi(u,\vec{y}) = \epsilon^{1/2} \Psi_1(u,\vec{y})+\epsilon^{3/2} \Psi_2(u,\vec{y})+\ldots, \label{eq:expan_Psi} \\ 
&A_{\mu}(u,\vec{y}) =  A_{\mu}^{(0)}(u,\vec{y})+\epsilon A_{\mu}^{(1)}(u,\vec{y})+\ldots, \label{eq:expan_A}
\end{align}
\end{subequations}
in which $\mu = (t,\phi,x,y)$. We assume the leading order contribution of the scalar field in the form of $\Psi_{1}(u,\vec{y})=\Phi(u,y)\text{e}^{i p x}$, with $p$ constant. Moreover, to have a constant magnetic field normal to the boundary, we shall consider $A_{x}^{(0)}(u,\vec{y})=B\,y$. In addition, a constant chemical potential at the boundary imposes $A_{t}^{(0)}(u,\vec{y})=A_{t}^{(0)}(u)$ and $A_{\phi}^{(0)}(u,\vec{y})=A_{\phi}^{(0)}(u)$. Thus, the nontrivial lowest-order Maxwell equations, coming from equation \eqref{eq:GenFieldEqns}, read
\begin{subequations}\label{eq:zeroth_maxwell}
	\begin{align} 
 {A_t^{(0)}}'' & -\Biggl[\qty(\Xi^2-1)\dfrac{f'}{f} -\dfrac{\Big(2(z-1)\Xi^2-z\Big)}{u}\Biggr]{A_{t}^{(0)}}'\nonumber\\
 & -\dfrac{a\Xi}{\ell^2}\Biggl[\dfrac{f'}{f}-\dfrac{2(z-1)}{u}\Biggr]{A_{\phi}^{(0)}}'=0 \label{eq:zeroth_maxwell0},\\ 
 {A_{\phi}^{(0)}}'' & +\Biggl[\Xi^2\dfrac{f'}{f} -\dfrac{\Big(2(z-1)\Xi^2+2-z\Big)}{u}\Biggr]{A_{\phi}^{(0)}}'\nonumber\\ & +a\Xi\Biggl[\dfrac{f'}{f}-\dfrac{2(z-1)}{u}\Biggr]{A_{t}^{(0)}}'=0,\label{eq:zeroth_maxwell1}
	\end{align}
\end{subequations}
while the remaining ones imply $A_{y}^{(0)}(u,\vec{y})=0$ and where the prime notation refers to derivatives with respect to $u$. At this order, the scalar and the Maxwell fields are not coupled, as expected from expansion \eqref{eq:epsilon_expan}. These two equations can be decoupled and analytically integrated. The following regular solutions arise
\begin{equation}\label{eq:zeroth_max_sols}
A_{t}^{(0)}(u) =  \mu - \rho \qty(\dfrac{u}{r_{h}})^{3-z},\quad  A_{\phi}^{(0)}(u) =  \nu- \zeta\qty(\dfrac{u}{r_{h}})^{3-z},
\end{equation}
with constant chemical potential $\mu$, constant charge density $\rho$ and $\zeta= -\dfrac{a}{\Xi}\, \rho$. Clearly, the above functions satisfy the required boundary conditions provided $1\leq z<3$. The regularity of the gauge field at the horizon translates to the vanishing of these components at $u=1$, requiring the fixing $\rho = r_{h}^{3-z}\mu$ and $\zeta = r_{h}^{3-z}\nu$. Notice that a non-trivial rotation parameter $a\neq0$ demands for a magnetic component of the gauge field. In the limit $a=0$ an electric gauge ansatz is sufficient to meet the boundary conditions. 

On the other hand, the lowest-order scalar equation 
has separable solutions of the form $\Phi(u,y) = R_n(u) \gamma_n(y;p)$, with the functions $R_n(u)$ and $\gamma_n(y;p)$ defined by the following equations
\begin{subequations}
\begin{align}
& \Bigl(-\partial^2_{YY}+\dfrac{Y^2}{4}\Bigr) \gamma_{n}(y;p) = \dfrac{\lambda_n}{2} \gamma_{n}(y;p), \label{eq:spatial}\\
R_{n}''(u) & +\Bigl(\dfrac{f'}{f}-\dfrac{z+2}{u}\Bigr) R_{n}'(u)=\Biggl[\qty(\dfrac{\ell^2}{r_h})^2 \dfrac{B \lambda_n}{f}\nonumber\\
& +\dfrac{\ell^2}{f}\Big(\dfrac{ m^2}{u^2}+\dfrac{(aA_t^{(0)}+\Xi A_\phi^{(0)})^2}{r_{h}^2}\Big)\nonumber\\
& -\qty(\dfrac{\ell u}{r_h})^{2(z-1)}\dfrac{(\ell^2\Xi A_t^{(0)}+aA_\phi^{(0)})^2}{r_{h}^2f^2}\Biggr]R_{n}(u)\label{eq:radial},
\end{align}
\end{subequations}
 where $Y:=\sqrt{2B}\qty(y-p/B)$. Equation \eqref{eq:spatial} is identified as a one-dimensional Schr\"odinger-like equation and is interpreted as the distribution of the order parameter on the $(x,y)$ plane.
Moreover, the radial equation \eqref{eq:radial} defines the superconducting phase transition. 

Let us now consider the regular and normalizable solutions of \eqref{eq:spatial}. These are spanned by the Hermite polynomials as 
 \begin{equation}\label{eq:droplet_sol}
    \gamma_{n}(y;p) =   \text{e}^{-Y^2/4} H_{n}(Y),
 \end{equation}
 with eigenvalues defined by $\lambda_{n}= 2n+1$, for a non-negative integer $n$. 
 
Now, we derive a general expression for the upper critical magnetic field $B_{c2}$ as a function of the background parameters. Near the boundary, the radial solution is given according to
\begin{equation}\label{eq:asym_rho}
R(u) = J_{-}u^{\Delta_{-}}+J_{+}u^{\Delta_{+}},		
\end{equation}
with $\Delta_{\pm}=\dfrac{z+3\pm\sqrt{(z+3)^2+4\ell^2m^2}}{2}$. To keep arguments as simple as possible, we also impose the boundary condition $J_{-}=0$, and set $J_{+}=J$ and $\Delta_{+}=\Delta$. 

The solution of \eqref{eq:radial} near the horizon is tackled through an expansion of $R(u)$ around $u=1$
\begin{equation}\label{eq:rho_expansion}
R(u) = R(1)+R'(1) (u-1)+\dfrac{1}{2}R''(1)(u-1)^2+\ldots\
\end{equation}
so that boundary conditions are met
\begin{subequations}
	\begin{align}
& R'(1)=-\dfrac{\ell^2 \qty(m^2 r_{h}^2+\ell^2 B)}{r_{h}^2 (z+3)}R(1),\\
& R''(1)= \dfrac{1}{2(z+3)^2}\Biggl\{-\qty(\dfrac{\ell}{\Xi})^2 \qty(\dfrac{\ell}{r_h})^{2z} [{A_t^{(0)}}'(1)]^2\\
&\,+\ell^2 m^2\Big[\ell^2 m^2+2(z+3)\Big]+\dfrac{\ell^{4} B}{r_{h}^2}\Big[2\ell^2 m^2+ \dfrac{\ell^4 B}{r_h^2}\Big] \Biggr\}R(1).	\nonumber
\end{align}
\end{subequations}
In what follows, we will use the semi-analytical matching method \cite{ZHAO2014438} which has shown good agreement with the numerical computations. This method is based on the mild assumption that there is an intermediate point $u=u_m$ where the horizon and boundary solutions, \eqref{eq:asym_rho} and \eqref{eq:rho_expansion}, as well as their first derivatives match. The resulting algebraic system can be solved consistently to give
\begin{equation}\label{eq:B_ext}
	B=\dfrac{r_{h}^2}{\ell^4 \eta}\Biggl\{ -\delta+\Biggl[\chi+\eta^2\qty(\dfrac{\ell}{\Xi})^2\qty(\dfrac{\ell}{r_{h}})^{2z} [{A_t^{(0)}}'(1)]^2\Biggr]^{1/2}\Biggr\},
\end{equation}
with 
\begin{align}
 \qquad\eta &=(1-u_m)\Bigl[(2-\Delta)u_m+\Delta\Bigr],\nonumber\\
 \quad\chi &= 2(z+3)\Big( 2(z+3)u_m^2-\ell^2m^2\eta^2\Big),\nonumber\\
\delta& = 2u_m\Bigl[(z+3)+\ell^2m^2(1-u_m) \Bigr]\nonumber\\
        & {}+\Delta (1-u_m)\Bigl[2(z+3)+\ell^2m^2(1-u_m) \Bigr] .
\end{align}
Near to the upper critical magnetic field value,  $B\approx B_{c2}$ is a good approximation. Hence, letting  \eqref{eq:zeroth_max_sols} in \eqref{eq:B_ext}, we disclose the expression
\begin{align}\label{eq:B_critical2}
	B_{c2}= & \qty(\dfrac{\ell^{z+1}\Xi}{z+3})^{2/z}\dfrac{T^{2/z}}{\ell^4 \eta}\Biggl\{ -\delta\nonumber\\
	& \qquad\qquad+\Biggl[\chi+\qty(\delta^2-\chi)\qty(\dfrac{T_{c}}{T})^{6/z}\Biggr]^{1/2}\Biggr\},
\end{align}
wherein the critical temperature is found to be
\begin{equation}\label{eq:T_critical}
    T_{c} = \dfrac{z+3}{\ell^{z+1} \Xi}\dfrac{1}{\qty(\delta^2-\chi)^{z/6}}\qty[\eta^2 \ell^{2z} \qty(z-3)^2 \qty(\dfrac{\ell}{\Xi})^2 \rho^2]^{z/6}.
\end{equation}
Moreover, it can be shown that when $T\approx T_c$ , the upper critical magnetic field has a linear dependence with the temperature
\begin{equation}\label{eq:B_critical_NTc}
    B_{c2} \propto \qty(1-\dfrac{T}{T_c}),
\end{equation}
in agreement with the well-known results in the Ginzburg-Landau theory of superconductivity.
%
%

\emph{Vortex lattice deformation.}---
\begin{figure*}
    \centering
    \begin{subfigure}[b]{.32\textwidth}
         \centering
         \includegraphics[width=\textwidth]{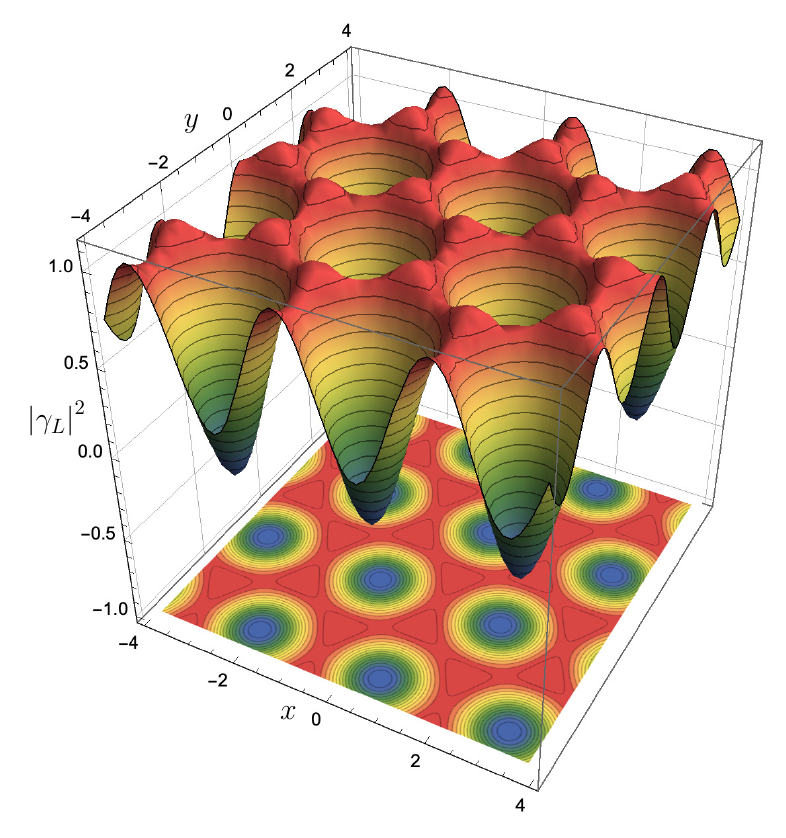}
         \caption{}
         \label{fig:sq_lat1}
    \end{subfigure}
    \begin{subfigure}[b]{.32\textwidth}
         \centering
         \includegraphics[width=\textwidth]{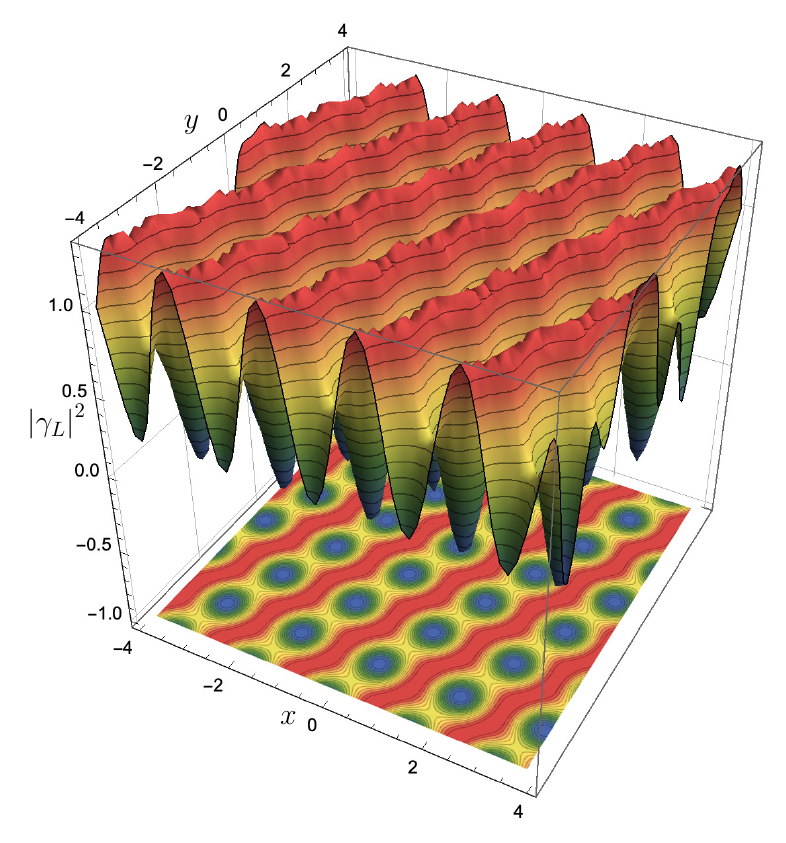}  
         \caption{}  
         \label{fig:rb_lat2}
       \end{subfigure}  
    \begin{subfigure}[b]{.32\textwidth}
         \centering
         \includegraphics[width=\textwidth]{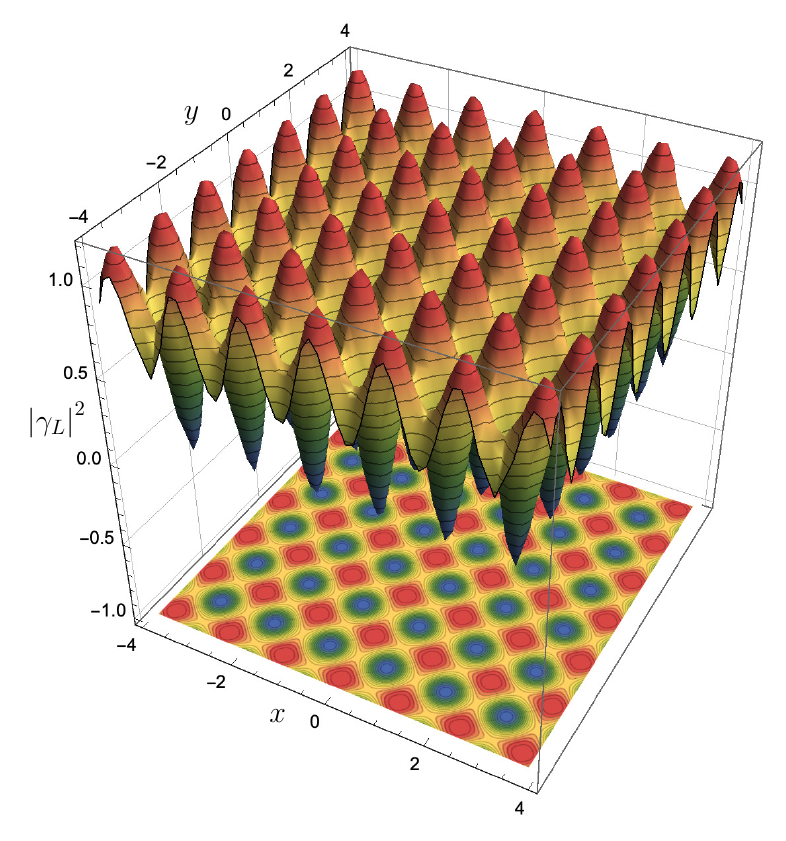}  
         \caption{}  
         \label{fig:tr_lat1}
    \end{subfigure}
        \caption{The transformation of the vortex lattice from triangular (a), passing through a rhombic form (b), to a square one (c), as the external magnetic field is increased. The vortex cores are located at the centers of the blue regions, wherein the function $\abs{\gamma_L}^2$ vanishes. The displayed figures depict a generic behavior independent of the choice of the background parameters. However, the critical dynamical exponent is restricted to $1\leq z<3$ due to renormalizability of the gauge field; see equation \eqref{eq:zeroth_max_sols}.} 
        \label{fig:lat_def2}
\end{figure*}
We are now in position to address the construction of the vortex lattice. We will follow a slightly similar procedure to the one implemented on \cite{Maeda:2009vf,Guo:2014wca}, by generalizing their symmetry arguments. We should note that it is sufficient to consider the lowest mode $n=0$ in the solution for $\gamma_{n}$ since it is expected to correspond with the most stable case \cite{Maeda:2009vf}. Namely, we take
\begin{equation}
\gamma_0(y;p_k) = \exp[-\dfrac{1}{2r_{0}^2}(y-p_kr_0^2)^2],
\end{equation}
with $r_0=1/\sqrt{B}$.
In this sense the lattice is constructed by a superposition of the lowest order solutions 
\begin{equation}\label{eq:lc_droplet}
\Psi_1(u,\vec{y}) = R_{0}(u) \sum_{k=-\infty}^{\infty}  c_{k} e^{ip_{k}x} \gamma_{0} (y;p_k),
\end{equation}
where $R_{0}(u)$ is the lowest order solution of the radial equation \eqref{eq:radial}.
At this point one finds useful to define the spatial factor in \eqref{eq:lc_droplet} as
\begin{equation}\label{eq:gamma_L}
    \gamma_{L}(\vec{y}) = \sum_{k=-\infty}^{\infty} c_{k} e^{ip_{k}x} \exp[-\dfrac{1}{2r_{0}^2}(y-p_k r_0^2)^2].
\end{equation}
This expression is very similar to the one emerging in Ginzburg-Landau theory for the order parameter of a type II superconductor under the influence of a constant magnetic field. The coherence length is identified as $r_0=\xi$, and close to the critical temperature it approaches
\begin{equation}\label{eq:B_xi_relation}
\xi \propto \qty(1-\dfrac{T}{T_{c}})^{-1/2},
\end{equation}
which also agrees with the known result in the Ginzburg-Landau theory. 

Note that the function \eqref{eq:gamma_L} is periodic along the $x$ direction but not necessarily along the $y$ direction. However, for this function to represent a vortex lattice, we need periodicity (at least up to a phase factor) along the $y$ direction as well. This is possible if we choose the parameters $c_k$, $q_{k}$, and $p_k$ to be
\begin{equation}\label{eq:parameters}
    c_{k}\equiv e^{iq_k},\qquad q_{k}\equiv \alpha k^2,\qquad p_{k}\equiv \beta k,
\end{equation}
with $\alpha$ and $\beta$ denoting arbitrary real constants. The above definition allows us to write $\gamma_{L}$ in terms of the elliptic theta function $\vartheta_3$ as 
\begin{equation}
    \gamma_{L}(\vec{y}) = e^{-y^2/2r_{0}^2} \vartheta_{3}(v,\tau),
\end{equation}
in which
\begin{equation}
    v = \dfrac{\beta}{2\pi}\qty(x-iy),\qquad \tau = \dfrac{1}{2\pi}\qty(2\alpha+i\beta^2r_{0}^2). 
\end{equation}
In this form, the symmetries of $\gamma_{L}$ are dictated by the well-known symmetries of $\vartheta_3$. Using this fact, we can easily show that $\gamma_{L}$ has a pseudo-periodicity in the following directions:
\begin{equation}\label{eq:fund_vectors}
\vec{b}_1 = \dfrac{2\pi}{\beta} \partial_x,\qquad\vec{b}_2=\dfrac{2\alpha}{\beta} \partial_x -\beta r_{0}^2 \partial_y.
\end{equation}
Here lies the keystone of the uttermost result. In contrast to \cite{Maeda:2009vf,Guo:2014wca}, we determine the most general form of \eqref{eq:fund_vectors}, which accommodates a lattice deformation beyond a rescaling by $r_0$. In consequence, the pseudo-periodicity of $\gamma_L$ reads
\begin{subequations}\label{eq:psi1_prop}
\begin{align}
 \gamma_{L}(x+\dfrac{2\pi}{\beta},y) &= \gamma_{L}(x,y) ,\\  
\gamma_{L}\Bigl(x+\dfrac{2\alpha}{\beta},y-\beta r_{0}^2\Bigr) &=\exp[-i\qty(\alpha +\beta x)]\gamma_{L}(x,y).
\end{align}
\end{subequations}
Altogether, the properties of $\gamma_{L}$ guarantee a lattice structure on $\abs{\gamma_L}^2$ with fundamental regions spanned by the vectors \eqref{eq:fund_vectors}. Therein, the magnetic flux penetrating a unit cell is quantized and given by $B\times 2\pi r_{0}^2 = 2\pi$. This coincides with the well-known results in the Ginzburg-Landau theory.
From \eqref{eq:fund_vectors}, we find the following relations between two adjacent vectors in the fundamental cell 
\begin{equation}\label{eq:vortex_pos}
 \cos{\theta}=\sqrt{1-\dfrac{\beta^4}{4\pi^2 B^2}},\qquad \alpha= \pi \cos{\theta},
\end{equation}
wherein $\theta$ denotes the angle between these two vectors and we have chosen their magnitudes to be the same.
Concerning the above relations, it is crucial to note that they provide a mechanism to continuously deform a given vortex lattice driven by the external magnetic field. Note that $\beta$ remains arbitrary in Eq.~\eqref{eq:vortex_pos}, hence it can be chosen appropriately as a function of the external magnetic field.

\emph{Experimental realization.}--- Beyond the theoretical scopes of the model here conceived, we present a direct application to the local vortex lattice deformations of a FeSe-based superconductor induced by an external magnetic field, as presented in \cite{Putilov:2019}. The experimental results show a transition from a nearly triangular lattice at $B=1$ T to a rhombic lattice at $B=3$ T, and an almost square one at $B=4$ T, at fixed temperature. Herein, T stands for the tesla unit, which is linked to the length scale in the bulk gravitational theory through T$=[\ell]^{-1}$. In our model, it suffices to determine an appropriate form of the $\beta$ parameter in terms of the external magnetic field. Our case of interest, the FeSe lattice of \cite{Putilov:2019}, is adequately comprised through the quadratic regression
\begin{equation}
 \beta = a_1+a_2\,B+a_3\,B^2, 
\end{equation} 
with the coefficients given by $a_1=2.37693\ \text{T}^{1/2}$, $a_2=-0.27870\ \text{T}^{-1/2}$ and $a_3=0.23444\ \text{T}^{-3/2}$.\\
 Fig.~\ref{fig:lat_def2} illustrates the evolution of a vortex structure when the external magnetic field increases. The observed configurations take place when the magnetic field adopts the values: $B_{tr}= 1$ T for the triangular structure, $B_{rh}=3$ T for the rhombic, and $B_{sq}=4$ T for the square one. Besides that, the vortex density increases when increasing the magnetic field, which also is in agreement with the experimental observations of \cite{Putilov:2019}. 
 
On the other hand, if we consider $\beta$ as a nonzero arbitrary constant, the vortex deformation will take place from a triangular to a square lattice, but now occurs by decreasing the external magnetic field. In this case, the triangular lattice corresponds to $B_{tr}=\beta^2/{\sqrt{3}\pi}$, while the square one corresponds to $B_{sq} = \beta^2/{2\pi}$, so that the transition is complete when the magnetic field has decreased about $13.4\%$ with respect to $B_{tr}$. However, to the best of our knowledge, this behavior has not been observed in experiments using superconducting materials. 

Finally, we advert to a less interesting case in which $\beta \sim \sqrt{B}$. In this particular setting, the external magnetic field has no other effect but scaling the vortex structure, similar to already known models. 


\emph{Conclusions and discussion.}---
In this work, we have constructed a holographic superconductor stemmed from a rotating Lifshitz background that mimics a type-II s-wave superconductor.

In constructing the Abrikosov vortex lattice, a generalized approach to that in \cite{Maeda:2009vf,Guo:2014wca} proved fruitful. We managed to describe continuous transformations of the vortex lattice by varying the magnetic field as a prediction of the holographic model. In particular, with a proper election of the parameter $\beta$ as a function of the external magnetic field, we faithfully reproduced the experimentally observed vortex transformation of the FeSe-based superconductor investigated in \cite{Putilov:2019}. Likewise, the transition of the lattice morphology reported in \cite{PhysRevB.99.161103} for a LiFeAs superconductor can also be modeled by our framework. Yet, the freedom in choosing $\beta$ means a further modeling power, worth investigating in detail with application to other phenomenological results.

An interesting point of comparison with our findings is provided by \cite{Xia:2021jzh}. There, the authors propose a numerical model capable of predicting the triangular and rectangular cases, at different temperatures. However, no other lattice configurations are reported, nor is there a mechanism for continuous deformations.

Finally, we emphasize that though our model was shaped from a near $B_{c2}$ approximation, the theory accurately describes transitions in the vortex lattice beyond the critical point. Thus, this work also paves the way to reproducing experimental magnetization curves by adjusting the free parameter $\beta(B)$. 

Besides, this holographic approach can be extended to other studies, namely, to describe holographic ferromagnetic or ferroelectric materials, and to analyze the magnetic response of materials simultaneously exhibiting both superconductivity and ferromagnetism.  

\begin{acknowledgments}
 All the authors are grateful to Manuel de la Cruz for enriching discussions. The authors acknowledge financial support from a CONACYT Grant No. A1-S-38041 and from the VIEP-BUAP Grant No. 122. JAHM also acknowledges support from CONACYT through a PhD Grant No. 750974. DFHB and JAMZ are also grateful to CONACYT for a \emph{Estancias posdoctorales por M\'{e}xico} Grant No. 372516 and project 898686, respectively. 
\end{acknowledgments}

\bibliography{Bibliography}

\end{document}